%% file: main.tex
\documentclass[sigconf,screen]{acmart}
\usepackage{xspace}
\usepackage{longtable}
\usepackage{array}
\usepackage[table]{xcolor}

\usepackage{color}
\usepackage{nameref}
\usepackage{multirow}
\usepackage{tabularx} 
\usepackage{censor}
\usepackage{nameref}
\usepackage{enumitem}

\usepackage[most]{tcolorbox}

\usepackage{xcolor}          

\definecolor{lightblue}{RGB}{0, 0, 100}

\newtcolorbox{MyBox}{
  colback=white,
  colframe=lightblue,
  fonttitle=\bfseries,
  coltitle=black,
  sharp corners,
  boxrule=1pt,
  left=5pt,
  right=5pt,
  top=5pt,
  bottom=5pt,
  breakable
}

\AtBeginDocument{%
  }

\setcopyright{acmlicensed}
\copyrightyear{2018}
\acmYear{2018}
\acmDOI{XXXXXXX.XXXXXXX}
\acmBooktitle{Companion Proceedings of the 34th ACM Symposium on the Foundations of Software Engineering (FSE ’26), July 5 - 9, 2026 Montreal, Canada}
\acmISBN{978-1-4503-XXXX-X/2018/06}

\begin{document}

\title{On the Economic Implications of Diversity in Software Engineering}

\author{Sofia Tapias Montana}
\email{sofia.tapiasmontana@ucalgary.ca}
\affiliation{%
  \institution{University of Calgary}
  \city{Calgary}
  \state{Alberta}
  \country{Canada}
}

\author{Ronnie de Souza Santos}
\email{ronnie.desouzasantos@ucalgary.ca}
\affiliation{%
  \institution{University of Calgary}
  \city{Calgary}
  \state{Alberta}
  \country{Canada}
  }

\begin{abstract}
\input{abstract}
\end{abstract}




\maketitle

\input{introduction}

\input{method}

\input{results}

\input{discussion}
\input{conclusion}

\nocite{*}

\bibliographystyle{ACM-Reference-Format}
\bibliography{bibliography}

\appendix

\end{document}

%% file: abstract.tex
This paper investigates how software professionals perceive the economic implications of diversity in software engineering teams. Motivated by a gap in software engineering research, which has largely emphasized socio technical and process related outcomes, we adopted a qualitative interview approach to capture practitioners’ reasoning about diversity in relation to economic and market oriented considerations. Based on interviews with ten software professionals, our analysis indicates that diversity is perceived as economically relevant through its associations with cost reduction and containment, revenue generation, time to market, process efficiency, innovation, and market alignment. Participants typically grounded these perceptions in concrete project experiences rather than abstract economic reasoning, framing diversity as a practical resource that supports project delivery, competitiveness, and organizational viability. Our findings provide preliminary empirical insights into how economic aspects of diversity are understood in software engineering practice.

%% file: introduction.tex
\section{Introduction}
\label{sec:introduction}

Equity, diversity, and inclusion have been recognized for several years as relevant concerns in software engineering, with sustained attention to the composition of software teams and its implications for software development work \cite{pieterse2006software, menezes2018diversity, albusays2021diversity}. Across this body of work, diversity is treated as an important component of software engineering practice and research, motivated by its association with collaboration quality, creativity, learning, innovation, and team effectiveness \cite{menezes2018diversity, rodriguez2021perceived, hyrynsalmi2025making}. Many studies characterize diversity through demographic and experiential attributes such as gender, age, nationality, culture, education, role, and personality, reflecting dimensions that are commonly used to describe software team composition \cite{pieterse2006software, menezes2018diversity, verwijs2023double}. The effects of diversity in software engineering are primarily reported in relation to socio-technical outcomes such as coordination, information sharing, psychological safety, conflict, and perceived effectiveness \cite{albusays2021diversity, devathasan2025empathy, hyrynsalmi2025making}, rather than through economic or delivery-oriented indicators.

Meanwhile, research in other fields such as management, information systems, and organizational communication places EDI more centrally within discussions of organizational performance and long-term project success \cite{arunachalam1995edi, jones2001user, okoro2012workforce, patricio2022systematic}. In these literatures, diversity is commonly framed as a component of human and intellectual capital and is associated with productivity, organizational effectiveness, innovation capacity, and competitive positioning \cite{okoro2012workforce, patricio2022systematic}. Although project-level financial outcomes are not always measured directly, economic and market-related considerations are treated as a primary motivation for EDI initiatives, with diversity and inclusion repeatedly associated with organizational performance, productivity, competitiveness, and, in some cases, financial outperformance at the firm level \cite{okoro2012workforce, patricio2022systematic, arunachalam1995edi, jones2001user, carucci2024one}. As a result, EDI is discussed not only as a social or workforce concern but also as a factor relevant to organizational viability and market competitiveness \cite{arunachalam1995edi, jones2001user, patricio2022systematic}.

This contrast points to a persistent gap in software engineering research regarding the economic, financial, and market-related implications of EDI. While SE studies provide detailed insights into how diversity shapes team processes and socio-technical dynamics \cite{albusays2021diversity, rodriguez2021perceived, verwijs2023double}, they rarely investigate how these effects translate into software project delivery outcomes, cost structures, commercial success, or industry-level performance \cite{albusays2021diversity, rodriguez2021perceived, verwijs2023double}. Compared to other fields where diversity is more explicitly connected to productivity, competitiveness, and organizational performance \cite{okoro2012workforce, patricio2022systematic, arunachalam1995edi, jones2001user, carucci2024one}, software engineering research remains largely focused on process-level and perceptual outcomes. This gap suggests that, although diversity and EDI are well established as important concerns within software engineering, their economic and market implications for software projects and the software industry remain underexplored.

Motivated by this gap, and by the increasing backlash and contestation surrounding equity, diversity, and inclusion initiatives in software engineering \cite{de2025diverse}, this research investigates diversity from an economic perspective rather than from predominantly socio-technical or process-oriented viewpoints \cite{albusays2021diversity, hyrynsalmi2025making, verwijs2023double}. While prior work in software engineering has emphasized collaboration, learning, and team dynamics, we deliberately shift focus toward how diversity is perceived to affect economic and market-related aspects of software work \cite{menezes2018diversity, rodriguez2021perceived, albusays2021diversity}. In particular, we investigate how software professionals understand the relationship between diversity in software teams and economic outcomes such as productivity, project success, and organizational value creation \cite{rodriguez2021perceived, okoro2012workforce, patricio2022systematic}. Based on this focus, we pose the research question \textbf{RQ. \textit{How do software professionals perceive the economic impacts of diversity in software teams?}} This paper reports initial results from a pilot qualitative study using interview data collected from software professionals, capturing how participants articulate and reason about these economic considerations in their work contexts.

%% file: method.tex
\section{Method}
\label{sec:method}

This study investigates how software professionals perceive the
economic implications of diversity in software engineering teams. Given the exploratory nature of
In this work and the limited prior work, we adopted a
qualitative interview design
\cite{seaman1999qualitative, hove2005experiences, ralph2020empirical}. The study was
organized into four sequential phases, as presented below.

\noindent \textbf{Participant Definition and Sampling.} Participants were software professionals with experience working
in different software development team configurations and professional contexts. Selection criteria focused on
ensuring variation in participants’ roles, organizational contexts,
and experiences. These criteria are consistent
with diversity and sampling practices recommended for qualitative
empirical research in software engineering
\cite{seaman1999qualitative,ralph2020empirical,lenberg2024qualitative}. Participants were recruited using a combination of convenience
sampling and snowball sampling \cite{baltes2022sampling}. Initial participants were identified through the researchers’
professional networks, and subsequent participants were recruited
through referrals from earlier interviewees. This strategy was
selected to facilitate access to practitioners willing to discuss
equity, diversity, and inclusion-related topics, which are currently
sensitive in professional settings \cite{de2025diverse}. The final sample consisted of 10 software professionals. Prior
qualitative studies in other fields report that thematic saturation
can be achieved with relatively small samples when the research
focus is narrow and exploratory \cite{guest2006many}. In this study, interviews were conducted until themes related to economic considerations of diversity began to recur. Given the pilot nature of the study, the sample is not intended to support statistical generalization, but rather to provide preliminary insights that can inform future, larger empirical investigations.

\noindent \textbf{Instrument Design and Piloting}
Data were collected through interviews supported by a structured interview guide \cite{hove2005experiences,seaman1999qualitative}. The guide was designed to elicit participants’ experiences and interpretations of diversity in software engineering teams, with particular attention to how these experiences were connected to economic and financial considerations in software projects. Questions addressed how participants described team diversity and team dynamics, how diversity was perceived to influence innovation, intellectual capital, and customer satisfaction, and how these factors were understood to relate to project-level financial gains or losses. Additional questions invited participants to compare diverse and homogeneous teams in terms of software outcomes and financial performance, and to reflect on the characteristics of teams perceived as well-positioned to deliver both high-quality software and favorable economic outcomes. Two pilot interviews were conducted before the main study to assess question wording, ordering, and flow. Insights from these pilot interviews informed minor adjustments to the script, improving readability and reducing redundancy, while preserving the substantive focus of the questions \cite{hove2005experiences, ralph2020empirical}.

\noindent \textbf{Data Collection Procedures.} Interviews were conducted individually using an online
videoconferencing platform. Each interview lasted between 25 and 60 minutes. Prior to each session, participants were
informed about the study’s purpose, the voluntary nature of
participation, and the measures taken to ensure confidentiality
\cite{ralph2020empirical,lenberg2024qualitative}. Written consent
for participation and recording was obtained from all participants. All interviews were recorded and subsequently transcribed for
analysis. Transcripts were reviewed for accuracy against the
recordings. The resulting dataset constitutes a preliminary corpus suitable for exploratory thematic
analysis.

\noindent \textbf{Data Analysis.} The interview data were analyzed using thematic analysis \cite{cruzes2011recommended}. Analysis began with repeated reading of the transcripts to develop familiarity with participants’ accounts and to gain an overall sense of how diversity was discussed in relation to a range of project-level and organizational considerations, including innovation, customer satisfaction, intellectual capital, and economic outcomes. Although the interviews elicited reflections across these dimensions, the analytic focus of this emergent results study was deliberately restricted. Specifically, our coding and analysis concentrated only on accounts in which participants directly and explicitly connected diversity to economic or financial aspects of software projects, such as costs, revenues, savings, or financial performance. Relevant text segments meeting this criterion were identified and labeled with descriptive codes capturing participants’ reasoning about diversity and financial outcomes. These codes were then examined for conceptual similarities and grouped into broader categories reflecting recurring patterns across interviews. Finally, categories were synthesized into themes that characterized how participants perceive the economic implications of diversity in software engineering teams. The analysis was iterative, involving movement between transcripts, codes, categories, and themes to ensure coherence and grounding in the data. Verbatim quotations were selected to support the reported themes while ensuring that no identifying or sensitive information was disclosed.

\noindent \textbf{Ethical Considerations.} The study was approved by the research ethics committee of the authors’ university. Participants were informed about the study objectives, data handling procedures, and confidentiality measures prior to participation. Given the sensitivity of equity, diversity, and inclusion-related topics \cite{de2025diverse}, full interview transcripts cannot be shared. Access to raw data is restricted to the research team, and only anonymized and non-sensitive quotations are reported in this paper to support the findings. All procedures adhered to established ethical practices for qualitative research in software engineering, including informed consent, protection of participant anonymity, and careful handling of potentially sensitive data \cite{ralph2020empirical,lenberg2024qualitative}.

%% file: results.tex
\section{Findings} \label{sec:findings}

Our study involved ten software professionals with diverse roles and levels of seniority within the software industry. Our sample included four participants in managerial or coordination positions, three participants in senior technical leadership roles related to software engineering, architecture, integration, or innovation, two senior individual contributors working as software engineers or technical architects, and one participant in an executive-level role with responsibility for organizational strategy and business development. With respect to seniority, six participants occupied senior positions, one participant was in a junior role, and the remaining three held mid to senior positions with substantial responsibility. Our sample was balanced in terms of gender, with five women and five men. Participants had experience working in a variety of software development contexts, including product development, platform and integration projects, innovation initiatives, and client-facing software delivery, often within international and cross-functional teams. This combination of technical, managerial, and strategic perspectives positioned our participants well to reflect on how diversity in software teams is perceived to relate to economic, financial, and market-oriented aspects of software projects.

\subsection{Perceived Economic Impacts of Diversity in Software Projects}
Participants articulated multiple ways in which diversity in software teams was associated with economic and financial aspects of software projects and organizations. These accounts varied in specificity and emphasis, ranging from direct references to costs, revenues, and savings to more indirect reasoning through market performance, efficiency, and innovation. The effects described by participants clustered around a set of recurring economic impact types, which are presented below.

\noindent \textbf{Cost Reduction and Cost Containment.}
Several participants explicitly associated diversity in software teams with reduced operational costs or the ability to achieve project objectives without increasing expenditures. These accounts emphasized how diverse perspectives enabled teams to anticipate problems, avoid rework, or pursue alternative technical solutions. One participant explained that what initially appeared to be a constrained technical situation evolved into a different solution that \textit{``reduced the cost of the operation'' (P03)}. Another participant described how working with people from different backgrounds and time zones allowed the team to meet project goals \textit{``without increasing costs, despite the delay we were already carrying'' (P03)}. In some cases, cost-related impacts were articulated in concrete monetary terms. For example, one participant reported that a diverse team was able to replace long-standing competitor software, noting that \textit{``we saved the company four million dollars a year that we were paying to our competitor'' (P05)}.

\noindent \textbf{Revenue Generation and Financial Gains.}
Participants also related diversity to revenue generation and broader financial performance in software projects. These accounts often framed diversity as enabling the delivery of software products that were more competitive or financially attractive in the market. One participant linked diversity to time-to-market considerations, stating that \textit{``if we look at financial gains in terms of time to market… having diversity allows us to arrive with a product that is more financially attractive to the target population'' (P01)}. The same participant further emphasized that diversity supported the achievement of \textit{``the financial results that are needed'' (P01)}. Other participants provided more explicit revenue-related examples. One participant explained that avoiding late-stage adjustments resulted in higher income, observing that \textit{``when those adjustments do not occur, revenues are much higher… we are easily talking about between one and two million dollars'' (P02)}. Similarly, another participant summarized this relationship by stating that \textit{``diversity produces innovation, innovation produces efficiency… and that translates into better sales'' (P02)}.

\noindent \textbf{Market Performance and Customer-Related Outcomes.}
In addition to direct financial outcomes, participants described diversity as shaping economic performance through market-facing mechanisms. Rather than focusing on immediate monetary metrics, these accounts emphasized product relevance, customer acquisition, and usability. One participant noted that diversity enabled teams to deliver software that was \textit{``more relevant… to the target population'' (P01)}. Another participant highlighted how diverse professional backgrounds contributed to usability, suggesting that this perspective \textit{``probably contributed to getting more customers than we expected'' (P06)}. Similarly, one participant emphasized that having a diverse team provided insight into different user groups, explaining that \textit{``having a diverse team gives you direct insight into the customer… and that firsthand insight helps facilitate that relationship'' (P08)}.

\noindent \textbf{Time-to-Market and Process Efficiency.}
Participants further associated diversity with time-to-market and process efficiency, which they framed as economically meaningful in software development. These accounts emphasized reduced delays, faster delivery, and avoidance of downstream corrections with financial implications. One participant explicitly framed efficiency in economic terms, stating, \textit{``if we look at financial gains in terms of time to market'' (P01)}. Another participant explained that minimizing adjustments had financial consequences, noting that \textit{``when those adjustments do not occur, revenues are much higher'' (P02)}. In other cases, efficiency was discussed in relation to competitive positioning, with one participant describing how a diverse team matched a competitor’s long-established software within a relatively short period, characterizing it as \textit{``an amazing turnaround'' (P05)}.

\noindent \textbf{Innovation and Solution Quality as Economic Enablers.}
Finally, participants described innovation and solution quality as mechanisms through which diversity enabled economic outcomes in software projects. Innovation was not framed as an economic outcome in itself, but as a condition that supported efficiency, market performance, or financial gains. One participant stated, \textit{``diversity produces innovation, innovation produces efficiency'' (P02)}, explicitly linking creative processes to economic consequences. Another participant emphasized that exposure to multiple viewpoints enabled the introduction of new software features, observing that \textit{``the more ideas bounce around, the more new things you can put into your software'' (P05)}. Some participants associated diversity with improved robustness and anticipation of user needs, noting that diverse teams \textit{``can look at all the different types of end users for a product and try to see the bigger picture'' (P07)}.

\subsection{Diversity, Intersectionality, and Economic Considerations.}
Across the interviews, participants occasionally referred to multiple dimensions of difference simultaneously, such as nationality, professional background, experience, or role, when describing software team composition and collaboration. These accounts point to an implicit and practice-oriented form of intersectionality, in which diversity is understood as the coexistence of multiple attributes rather than as a single categorical dimension. Participants did not, however, frame these combinations in terms of intersecting systems of advantage or disadvantage, nor did they use intersectionality as an explanatory concept in their reasoning. When economic or financial aspects were discussed, participants’ reasoning remained largely aggregated at the team or project level rather than differentiated by intersecting identities. In some cases, references to multiple backgrounds or perspectives were linked to improved customer understanding, usability, or problem anticipation, which participants treated as economically consequential within the software industry context. For example, one participant emphasized that teams with \textit{``different backgrounds, maybe different nationalities''} needed to attend closely to customers, because \textit{``that is where a company’s financial gain comes from''} (P06). Similarly, another participant noted that having \textit{``different perspectives from the team''} enabled insight into \textit{``diverse customers''}, facilitating economically relevant customer relationships (P08).

\subsection{Answering RQ: How do software professionals perceive the economic impacts of diversity in software teams?}
Software professionals perceive the economic impacts of diversity in software teams in varied and context-specific ways, drawing primarily on concrete project experiences rather than abstract economic reasoning. Participants associated diversity with economic aspects such as cost reduction and cost containment, revenue generation, and market performance, often grounding these perceptions in situations involving alternative technical solutions, avoidance of rework, improved product relevance, or, in some cases, quantified financial savings. Participants also described diversity as economically relevant through its influence on time-to-market, process efficiency, and innovation, which were consistently framed as enabling conditions that support financial outcomes rather than as economic outcomes in themselves. In several accounts, economic relevance was articulated indirectly, through improved customer understanding, usability, or alignment with target users, which participants treated as consequential for the financial performance of software products. Overall, software professionals framed the economic significance of diversity in terms of how it supports the delivery, competitiveness, and viability of software projects within their specific organizational and project contexts.

%% file: discussion.tex
\section{Discussion} \label{sec:discussion}

\noindent \textbf{Comparison with the Literature.} Prior software engineering research on diversity has primarily examined socio technical and process oriented outcomes, including coordination, communication, psychological safety, creativity, learning, and perceived effectiveness \cite{albusays2021diversity, menezes2018diversity, rodriguez2021perceived, verwijs2023double, devathasan2025empathy, hyrynsalmi2025making}. Our findings are aligned with this literature at the level of mechanisms. Participants described diversity through improved problem framing, anticipation of issues, solution quality, and collaborative reasoning, which resemble patterns reported in earlier software engineering studies \cite{albusays2021diversity, menezes2018diversity, rodriguez2021perceived, verwijs2023double, hyrynsalmi2025making}. However, our study differs in analytical focus. Because participants were explicitly asked to reflect on economic and financial aspects, these socio-technical mechanisms were articulated as means through which diversity was perceived to contribute to economic and financial outcomes, including cost containment, revenue generation, time to market, and project viability. This indicates continuity with prior software engineering work in how diversity operates, while extending it by connecting these operations to economic reasoning that is largely absent from existing studies. When compared with the literature outside software engineering, our findings are closer in substance and framing to research in management, organizational communication, and information systems, where diversity is commonly treated as a component of human and intellectual capital that supports productivity, innovation, competitiveness, and value creation \cite{arunachalam1995edi, okoro2012workforce, patricio2022systematic, jones2001user, horwitz2007effects, gerpott2015differences}. At the same time, our findings remain distinct in how these economic links are articulated. Rather than relying on abstract performance indicators or firm-level measures, participants grounded economic reasoning in concrete software project experiences, technical decisions, and delivery constraints \cite{carucci2024one}. Overall, our study improves understanding of diversity in software engineering by showing that software professionals perceive diversity in software teams not only as a socio-technical condition, but also as an active mechanism through which economic and financial outcomes are pursued in practice, thereby strengthening connections between software engineering research and adjacent economic perspectives on diversity. \\

\noindent \textbf{Implications for Research and Practice.} From a research perspective, our study extends diversity research in software engineering by shifting attention toward a pragmatic and underexplored aspect of the software industry: economic and financial considerations. Prior work has predominantly focused on socio-technical processes, whereas our study foregrounds how diversity in software teams is perceived as a mechanism that contributes to economic and financial outcomes. By explicitly eliciting participants’ experiences about costs, revenues, efficiency, time to market, and market performance, our findings broaden the outcome space typically associated with diversity in software engineering. These findings suggest directions for more in-depth empirical studies that investigate how software team diversity and EDI practices relate to economic aspects of software development at project, organizational, and industry levels. Future research may build on this work to characterize conditions under which such economic impacts are achieved, how they interact with organizational context, and how they evolve over time. From a practice perspective, our findings have implications in a context where EDI initiatives in the software industry are increasingly contested. Our results provide preliminary evidence that diversity in software teams can contribute not only to social or ethical objectives but also to economic and financial results. Participants associated diversity with cost containment, revenue generation, efficiency, innovation, and market alignment, framing it as an element that supports the viability and competitiveness of software projects. This suggests that, for practitioners and decision makers, diversity can be understood as a practical asset in software development rather than solely as a normative commitment. In environments where EDI efforts are questioned or deprioritized, these findings offer an empirically grounded perspective indicating that diversity may yield tangible economic benefits for software teams and organizations. \\

\noindent \textbf{Threats to Validity.} As with any qualitative study, this study has limitations inherent to the research method. First, diversity was operationalized through participants’ self-descriptions and reflections on team composition rather than through standardized measures, which may have led to variability in how diversity was interpreted; to mitigate this, the interview guide explicitly prompted discussion of multiple diversity dimensions, supporting more consistent and comprehensive accounts across participants. Second, the study relied on a small, non probabilistic sample of ten software professionals recruited through convenience and snowball sampling, which limits transferability to other software engineering contexts; this was partially mitigated by recruiting participants from different roles and work settings, and by framing the findings as preliminary patterns that software companies and practitioners are expected to interpret and map onto their own contexts to assess fit. Third, the sensitive nature of equity, diversity, and inclusion topics may have influenced participants’ willingness to disclose negative or contested experiences; this risk was mitigated through ethics approval, clear assurances of anonymity and confidentiality, and neutral, non-judgmental question framing. Finally, as a pilot study, the analysis does not capture the full range of possible economic dimensions of diversity; this limitation was addressed by adopting an exploratory thematic analysis focused on recurring patterns supported by quotations, intended to inform subsequent studies with broader samples and complementary methods.

%% file: conclusion.tex
\section{Conclusions and Future Work} \label{sec:conclusions}

This study set out to investigate how software professionals perceive the economic implications of diversity in software engineering teams, with a deliberate focus on financial and market-related reasoning. Based on the software practitioners' experiences, our preliminary results suggest that diversity is perceived as influencing cost reduction and containment, revenue generation, time to market, process efficiency, and market alignment, often through mechanisms such as improved problem-solving, innovation, customer understanding, and avoidance of rework. These findings remain exploratory and are intended to inform subsequent empirical work rather than to support generalization. Our future work will extend this investigation beyond participants’ explicit accounts of finance and economy captured in the interviews, by exploring reported experiences related to innovation, customer satisfaction, and intellectual capital in the software industry.